\documentclass[graybox]{svmult}

\usepackage{aastexdefs}
\usepackage{natbib}

\usepackage{mathptmx}       
\usepackage{helvet}         
\usepackage{courier}        
\usepackage{type1cm}        
\usepackage{url}
\usepackage{makeidx}         
\usepackage{graphicx}        
\usepackage{multicol}        
\usepackage[bottom]{footmisc}

\usepackage{color}
\usepackage{amssymb}

\makeindex             

\begin{document}

\title*{Supernova Remnants as Clues to Their Progenitors}
\author{Daniel Patnaude and Carles Badenes}
\institute{Daniel Patnaude \at Smithsonian Astrophysical Observatory, 60 Garden St, Cambridge, MA 02138,
  \email{dpatnaude@cfa.harvard.edu} \and Carles Badenes \at Department of Physics and Astronomy, and Pittsburgh Particle Physics,
  Astrophysics, and Cosmology Center (PITT-PACC), University of Pittsburgh, 3931 O'Hara St, Pittsburgh, PA 15260
  \email{badenes@pitt.edu}}
%
%
\maketitle

\abstract{Supernovae shape the interstellar medium, chemically enrich their
host galaxies, and generate powerful interstellar shocks that drive future
generations of star formation. The shock produced by a supernova event acts as
a type of time machine, probing the mass loss history of the progenitor system
back to ages of $\sim$ 10 000 years before the explosion, whereas supernova
remnants probe a much earlier stage of stellar evolution, interacting with
material expelled during the progenitor's much earlier evolution.  In this
chapter we will review how observations of supernova remnants allow us to infer
fundamental properties of the progenitor system.  We will provide detailed
examples of how bulk characteristics of a remnant, such as its chemical
composition and dynamics, allow us to infer properties of the progenitor
evolution. In the latter half of this chapter, we will show how this exercise
may be extended from individual objects to SNR as classes of objects, and how
there are clear bifurcations in the dynamics and spectral characteristics of
core collapse and thermonuclear supernova remnants. We will finish the chapter
by touching on recent advances in the modeling of massive stars, and the
implications for observable properties of supernovae and their remnants.}

\section{Introduction}
\label{sec:intro}

Supernova remnants provide a unique view of the supernova (SN) phenomenon.
The hundreds of extragalactic supernovae (SNe) discovered each year by
modern surveys are unresolved point sources which fade before they can 
interact with any circumstellar material outside of their most immediate
surroundings. Before they fade, they probe the evolution of the progenitor
on timescales of $\sim$ 100--1000 years before the explosion. This is
because the typical outflow speeds from the progenitor are $\sim$ 
10--1000 km s$^{-1}$, 100-1000$\times$ slower than the typical
supernova blastwave velocity.

In contrast, supernova remnants (SNRs) in the Local Group galaxies can be 
resolved in some detail as they interact with the bulk of the circumstellar
material expelled by their progenitors during pre-SN mass loss over
much longer timescales. Additionally,
SNRs are surrounded by \textit{resolved} stellar populations that spawned
their SN progenitors \citep{badenes09,williams14,jennings14}, 
and they have become prime hunting grounds in the
search for their surviving binary companions 
\citep[e.g.,][]{edwards12,kerzendorf14}. X-ray spectra
of SNRs can reveal emission from neutron-rich isotopes in
the ejecta which are sensitive probes of SN explosion mechanisms
\citep[e.g.,][]{yamaguchi15}. In
short, SNR observations can constrain aspects of SNe and their 
progenitors that are either 
difficult or impossible to access by optical SN observations. 

In this chapter, we review the clues that supernova remnants provide about 
their progenitor's evolution and ultimate fate. We 
begin by highlighting studies of both Type Ia and core-collapse 
SNRs which inferred properties of their progenitors and 
evolution.  We then discuss recent techniques that use bulk SNR 
observables to constrain SN progenitor properties. We
conclude by touching on recent results in massive star 
evolution which may be relevant towards understanding
observable properties in supernova remnants. For further background and 
a different perspective of these topics, see the reviews
by \citet{badenes10} and \citet{vink12}.

\section{Supernova Remnant - Progenitor Connections}
\label{sec:overview}
\index{supernova remnants}
\index{progenitors}

The main challenge associated with studying supernova remnants is
related to their type. Classically, SNe have been typed by their optical 
spectrum around maximum light \citep{filippenko97}. 
In general terms, Type Ia supernovae show
no hydrogen in their optical spectra, with Type Ia (thermonuclear)
SNe exhibiting strong \ion{Si}{2} absorption around 615 nm. Type Ib/c
are core-collapse supernovae, and differ from Ia SNe in that they possess
either a weak or non-existent silicon absorption feature, and 
little (Type Ib) or no (Ic) absorption from \ion{He}{1}. The amount
of helium in the optical spectrum may provide insight into the degree
of mass-loss in the progenitor prior to core-collapse. 
In contrast, Type II SNe do show hydrogen in their spectra, and
are further subdivided into types IIP, IIL, IIb and IIn, 
based on the evolution and characteristics 
of their light curve or, in the case of Type IIn SNe, their optical
spectrum. 

While SNe are common and somewhat accessible to typing, SNRs are evolved
objects, and \textit{a priori} it is non-trivial to associate them with each of
the established SN types, though the detection of optical light echoes from the
SN event do provide clues in some cases \citep{rest08}. However, given the
wealth of data available, a challenging opportunity exists to connect remnant
properties back to the properties of the progenitor. 

Connecting remnant properties to those of the progenitor often
involves a detailed analysis of the remnant's broadband spectrum,
combined with comparisons to hydrodynamical models for supernova remnant ejecta
evolution. In some cases such as those discussed below, this has
produced interesting constraints. However, bulk properties such as emission
line strengths, energy centroids, and chemical abundances depend on
different aspects of a SN explosion than those of optical SNe. These properties
can act as crucial benchmarks for specific stellar evolution scenarios. In
\S~\ref{sec:bulk} of this chapter, we explore how these so called ``bulk
observables'' connect directly to the evolution of the progenitor, and how
theorists and observers are connecting detailed stellar evolution models to
properties of young supernova remnants.

When studying SNRs, both as individual objects and as a class, the main
properties to consider are the chemical composition of the ejecta and the
dynamics of the blast wave. The spectrum of a SNR in the X-ray band is a
combination of nonthermal emission from the blast wave, and thermal emission
from both swept-up circumstellar material and shocked supernova ejecta. The
chemical composition of the shocked ejecta is determined by fitting models to
the observed X-ray spectrum, which often requires some knowledge of the SNR
dynamics \citep{borkowski01}. In turn, the blast wave dynamics are inferred by
relying on critical assumptions about the relationship between the shock speed
and the measured electron temperature. Only in certain cases, such as in the
Galactic SNR Cas A, are the dynamics directly observed
\citep{delaney03,patnaude09}. 

In most cases, the morphology is compared to
self-similar models for blast wave evolution, which again require assumptions
about the remant's age, energetics, and circumstellar environment. These
methods are vulnerable to circular logic and can result in detailed statements
about the properties of a given SNR that fail to properly account for the
considerable uncertainties introduced by the models. In other words, the
critical assumptions made in connecting spectral properties to dynamics often
compromise the the connections made back to the progenitor evolution.

\begin{figure}
\includegraphics[width=12cm]{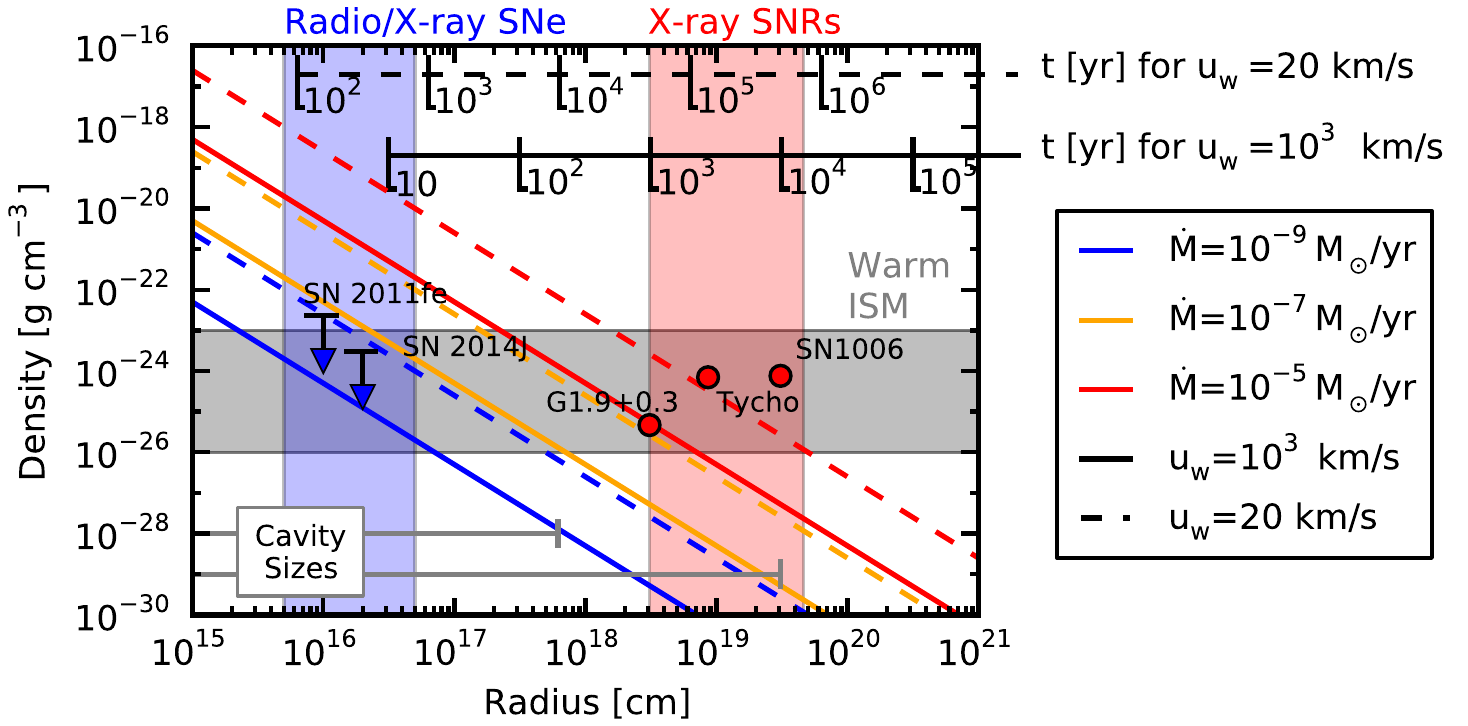}
\caption{Schematic circumstellar density profiles for isotropic SN Ia progenitor outflows (colored plots, dashed for
  $\mathrm{u_{w}} = 20$ \kms; solid for $\mathrm{u_{w}} = 10^3$ \kms), compared to the typical densities in the warm phase of the
  interstellar medium (horizontal grey band) \citep{ferriere01}. The rulers on top of the plots show the time before the explosion
  (in yr) when the material at that radius was ejected by the SN progenitor. The rulers on the bottom show the typical sizes of
  circumstellar cavities excavated by fast outflows \citep{badenes07,shen13}. The radii probed by X-ray/radio follow-up
  observations of SNe and studies of Type Ia SNRs are shown by blue and red shaded vertical stripes.  The blue arrows and red
  circles show upper limits for two Type Ia SNe \citep{chomiuk16,perez14} and measurements for three Type Ia SNRs
  \citep{raymond07,slane14,borkowski14}. }
\label{fig:snr_r_rho}
\end{figure}

As an example, specific stellar evolution scenarios from SN Ia progenitors
predict circumstellar structures that range from the $\rho \propto r^{-2}$
profiles that are the signature of isotropic mass-loss to relatively large
low-density energy-driven cavities excavated by fast outflows from accreting
white dwarfs \citep{badenes07,shen13}. As seen in Figure~\ref{fig:snr_r_rho},
the prompt radio and X-ray emission from SNe \citep{chomiuk16,margutti14},
weeks or months after the explosion, can only probe circumstellar material
(CSM) structures up to a few hundred AU, composed of material
expelled by the progenitor hundreds to thousands of years before the explosion,
depending on the outflow velocity.  

In contrast, SNRs that are hundreds
or thousands of years old can probe CSM structures up to several pc in size,
expelled by the progenitor several thousands to millions of years before the
explosion. These spatial scales are representative of typical stellar wind
blown bubbles \citep{koo92}, and the associated temporal scales are directly
relevant to stellar evolution scenarios for Type Ia SN progenitors
\citep{wang12}. For CC SNRs, the argument is even stronger, as mass-loss is
often the single most uncertain ingredient in stellar evolution models for SN
progenitors \citep{smith14}.

A different approach is instead of studying the minute
details of individual objects, 
one probes the general characteristics of SNRs to
see how they relate to the expected characteristics of a model derived
from a potential progenitor scenario. This approach has been 
successful in Type Ia SNRs, and is now beginning to be explored in
core collapse scenarios.

\subsection{Type Ia SNR -- Progenitor Connections}
\label{sec:typeIa_sne}
\index{Type Ia SNR}

Despite decades of efforts, the progenitors of Type Ia SNe remain unidentified.
The absence of hydrogen and helium in the optical spectra of SN~Ia, their
association with both young and old stellar populations, the absence of a
detectable shock breakout in the early light curves, and the chemical
composition of the ejecta, as inferred from optical and X-ray observations,
imply that the exploding star is a carbon-oxygen (C/O) white dwarf (WD). The
exact events leading to the explosion are unclear, but after the degenerate
matter is somehow ignited, a thermonuclear runaway consumes most of the star,
synthesizing 0.5-1.0M$_{\sun}$ of $^{56}$Ni, which powers the optical
lightcurve through the decay chain $^{56}$Ni $\rightarrow$ $^{56}$Co
$\rightarrow$ $^{56}$Fe. 

Different SN Ia progenitor scenarios attempt to provide a physical explanation
for the destabilization of the WD and the subsequent explosion. In the
single-degenerate (SD) scenario, the WD grows slowly in mass by accretion from
a non-degenerate companion, until it gets close to the Chandrasekhar limit
(M$_{\mathrm{Ch}}$ $\approx$ 1.4M$_{\sun}$). In the double degenerate (DD)
scenario, the WD explodes after a dynamical merger or a collision with another
WD, forming a short-lived object whose mass does not necessarily have to be
M$_{\mathrm{Ch}}$. See \citet{hillebrandt13} and \citet{maoz14} for recent
reviews on SN~Ia progenitors.

The so-called `SN~Ia progenitor problem' persists because even though there is
a great deal of indirect evidence supporting both scenarios, a definite
`smoking gun' remains elusive \citep[see][for a discussion]{maoz14}. In the SD
scenario, the WD accretes mass at a rate of $\dot{M}$ = $\sim$10$^{-7}$
M$_{\sun}$ yr$^{-1}$ \citep{shen07}. At these rates, stable hydrogen burning
can occur on the surface of the WD, which allows steady mass growth to
M$_{\mathrm{Ch}}$.  Accretion at higher rates can lead to the expansion of the
accretor and engulfment of the companion into a common envelope phase, which
effectively halts the growth of the WD. Solutions to this problem have been
proposed, however, in the form of fast optically thick accretion winds that
regulate the accretion \citep{hachisu96}. Lower accretion rates can result in
the accumulation of degenerate material on the WD, which will eventually ignite
as a nova where much of the accreted material and potentially some of the
accretor mass is ejected, effectively eroding the mass of the WD. 

In the DD scenario, a more massive white dwarf
merges or collides with a lower-mass WD and this rapid episode of mass accretion
can effectively avoid problems related to inefficient mass growth in the SD
scenario, and result in a SN Ia explosion, not necessarily close to
M$_{\mathrm{Ch}}$ \citep{vanKerkwijk10}. However, off-center ignition in the
primary WD can result in the formation of a O/Ne WD and an accretion-induced
collapse to a neutron star without a SN Ia explosion \citep{saio85}. 

Thus two key observational probes of the SN Ia progenitor problem are the
total mass of the ejecta (i.e., whether it is close to M$_{\mathrm{Ch}}$ or
not) and the presence or absence of a fast accretion wind outflow before the
explosion. SNR studies can effectively constrain both ejecta
mass and pre-explosion outflows.

When a WD explodes close to M$_{\mathrm{Ch}}$ in the SD scenario, it will have
a dense ($\rho \gtrsim$ 2$\times$10$^{8}$ g cm$^{-3}$) core where efficient
electron captures can take place. This results in a significant production of
neutron-rich isotopes such as $^{58}$Ni and $^{55}$Mn in the innermost
$\sim$0.2 M$_{\sun}$ of the WD \citep{brachwitz00}. In contrast, a SN Ia
explosion with a DD progenitor will not present this neutronized core, unless
it happened exactly at M$_{\mathrm{Ch}}$.  

\begin{figure}
\includegraphics[clip=true,viewport=0 360 406 768,width=0.465\textwidth]{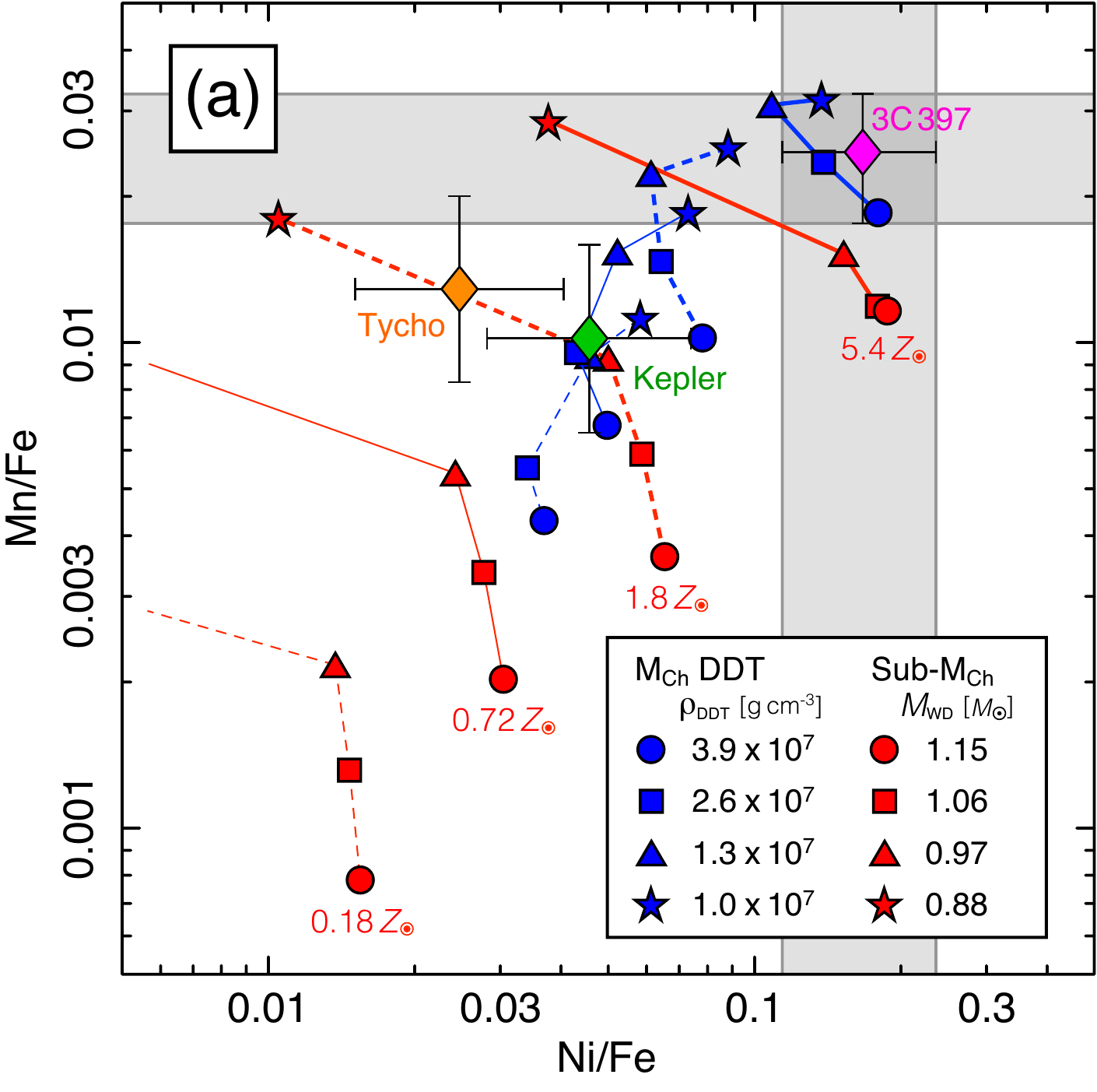}
\includegraphics[clip=true,viewport=0 0 406 352,width=0.535\textwidth]{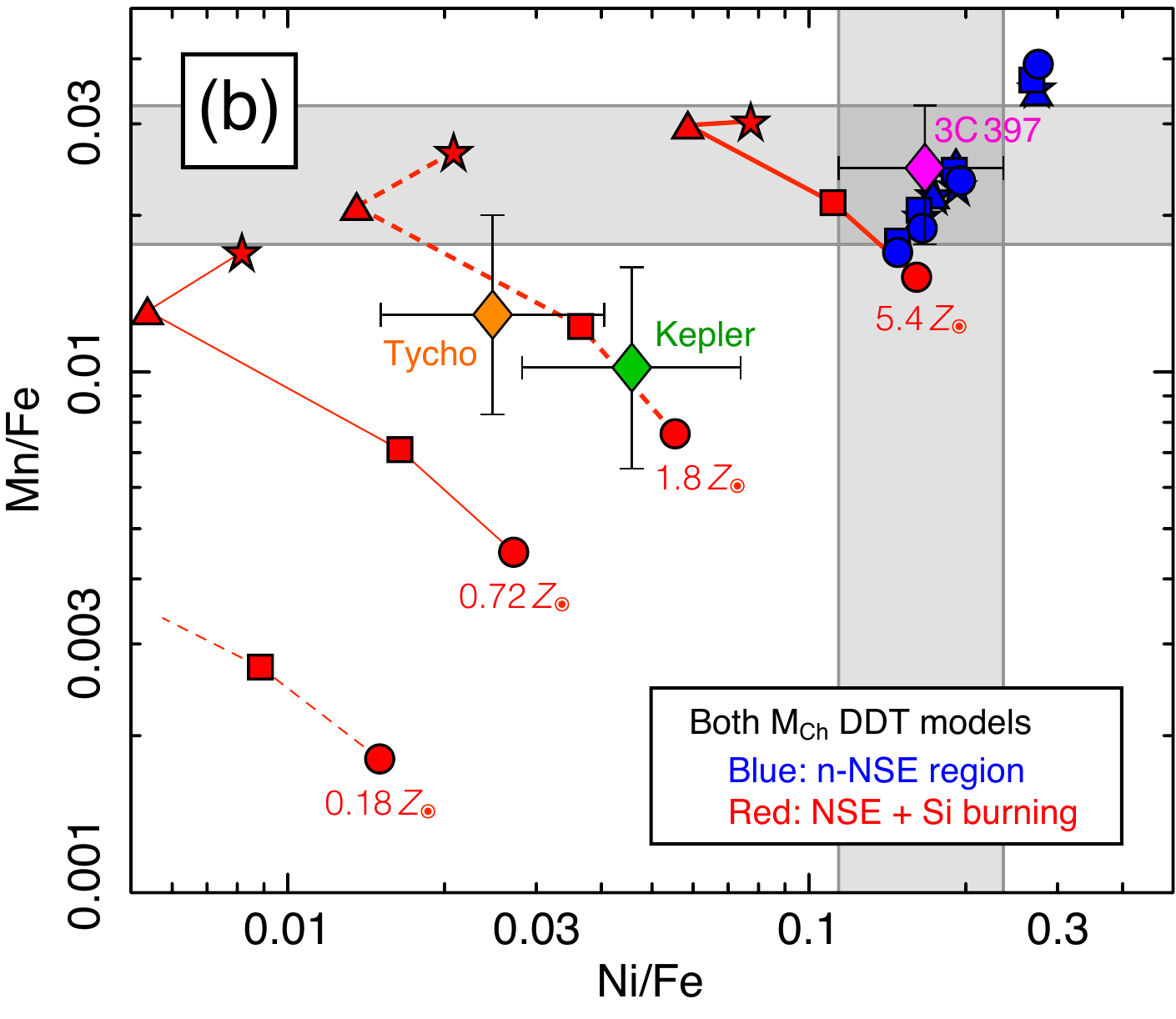}
\caption{(a) Ni/Fe and Mn/Fe mass ratios in SN Ia models, compared with the observed values in 3C~397 (magenta diamond and gray
  regions). Blue symbols represent $M_{\rm Ch}$ delayed-detonation models with different values of $\rho_{\rm DDT}$ as indicated
  in the legend. Red symbols represent sub-$M_{\rm Ch}$ detonation models with WD masses given in the legend. Models with the same
  progenitor metallicity are connected by lines: 5.4\,$Z_\sun$ (thick solid), 1.8\,$Z_\sun$ (thick dashed), 0.72\,$Z_\sun$ (thin
  solid), and 0.18\,$Z_\sun$ (thin dashed), The values for Kepler and Tycho indicated by the green and orange diamonds are
  calculated using the line fluxes from \citet{park13} and \citet{yamaguchi14}.  (b) Same as panel (a), but the values predicted for
  the innermost 0.2 M$_{\sun}$ that is dominated by the n-NSE regime (blue) and the other regions (red) from the $M_{\rm Ch}$
  models are separately shown. The observed mass ratios for 3C~397 can be well explained by the nucleosynthesis that occurs in the
  n-NSE regions of standard metallicity models. \citep[Figures reproduced with the
permission of the authors; ][ Figure 4, pp. 5]{yamaguchi15}.}
\label{fig:3c397}
\end{figure}

X-ray observations of Type Ia SNR
can reveal the presence or absence of this neutronized core by comparing
diagnostic mass ratios like Ni/Fe and Mn/Fe to the predictions of both
M$_{\mathrm{Ch}}$ and sub-M$_{\mathrm{Ch}}$ models \citep{park13}. In the
evolved Galactic SNR 3C~397, \citet{yamaguchi15} found high ratios of Ni/Fe and
Mn/Fe in the X-ray spectrum. As seen in Figure~\ref{fig:3c397}, the Ni/Fe and
Mn/Fe ratios are among the highest reported for Type Ia SNRs, and strongly
indicate that the progenitor must have exploded close to M$_{\mathrm{Ch}}$,
which is more naturally explained by the SD scenario.

A key prediction of the SD scenario is the presence of circumstellar material
(CSM). One possible source of this CSM are the fast, optically thick winds that
are required to stabilize accretion and allow the WD to grow to
M$_{\mathrm{Ch}}$ and avoid a common-envelope phase \citep{hachisu96}. These
fast winds have mass loss rates comparable to the accretion rate (1-5 $\times$
10$^{-7}$ M$_{\sun}$ yr$^{-1}$), and velocities larger than the escape velocity
from the surface of a massive WD (several hundred km s $^{-1}$), resulting in
large mechanical luminosities. 

By the time of the SN explosion, these outflows should have carved out large
(several pc in diameter) low-density cavities around the progenitor system.
Even if accretion winds are not active, it is hard to envision any realization
of the SD scenario that does not lead to some kind of CSM, because the
accretion process cannot be 100\% efficient. For outflow velocities slower than
those found in accretion winds, the CSM should stay close to the progenitor,
and geometric dilution will lead to a density profile of the form
$\rho_{\mathrm{CSM}}$ $\sim$ $Ar^{-2}$, where $A$ is a constant of
proportionality dependent upon the mass-loss rate and wind velocity.
\citet{badenes07} studied the impact of several CSM distributions on SNR
evolution, and concluded that most Type Ia SNRs show no significant evidence
for the presence of CSM. 

\begin{figure}
\centering
\includegraphics[width=0.5\textwidth]{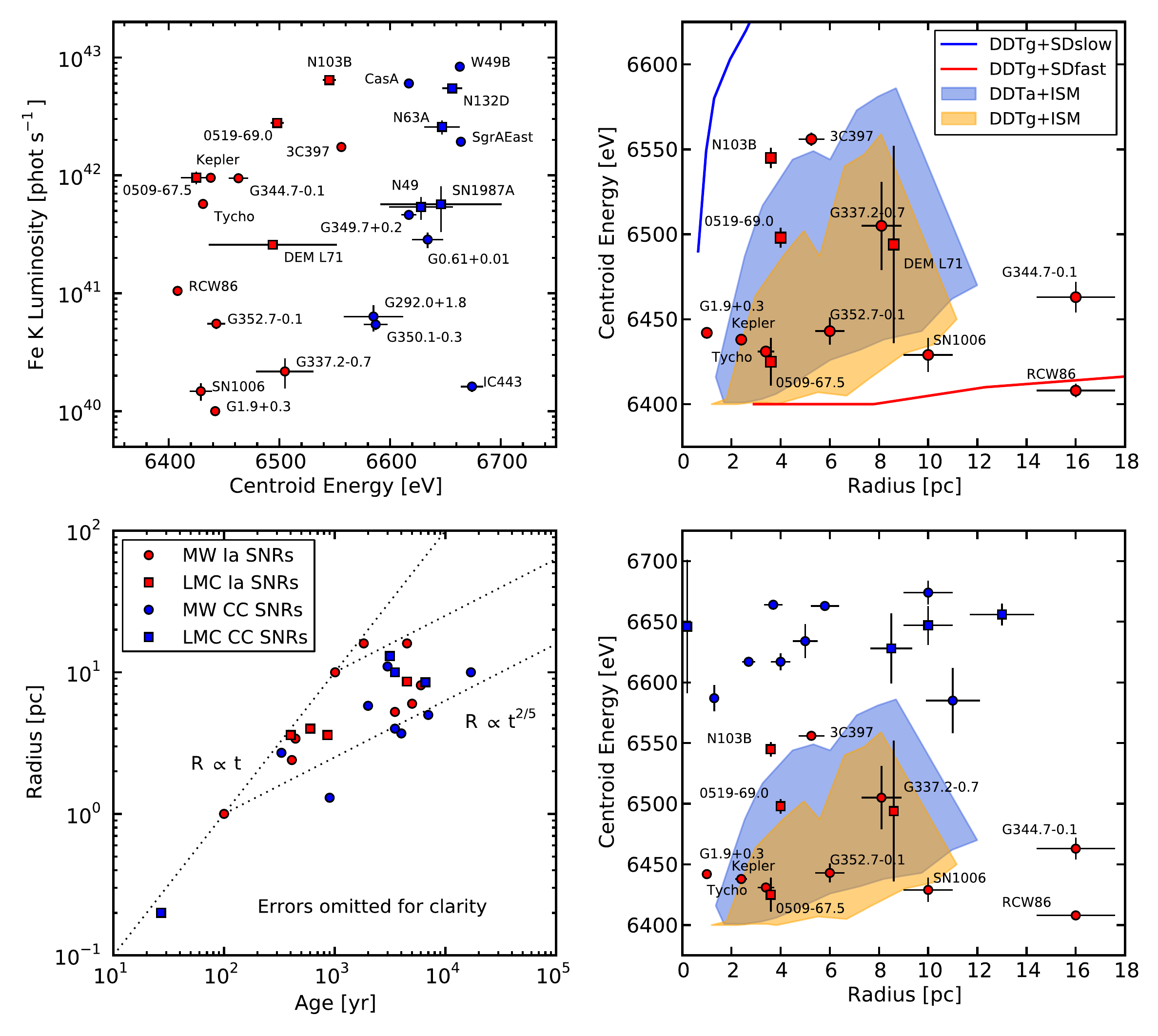}
\caption{Fe K$\alpha$ line centroid as a function of radius for the Type Ia SNRs in \citet{yamaguchi14}, plus DEM L71
  \citep{maggi16}. The shaded regions represent theoretical predictions from Type Ia SNR models interacting with a uniform
  interstellar medium \citep{badenes03,badenes05}: orange for the $^{56}$Ni-poor delayed detonation model DDTg, blue for bright
  $^{56}$Ni-rich delayed detonation model DDTa. The colored lines represent theoretical predictions from \citet{badenes07} for
  model DDTg interacting with two circumstellar profiles calculated assuming an SD progenitor, red for an accretion wind outflow
  with a large cavity (SDfast) and blue for a $\rho \propto r^{-2}$ outflow (SDslow).}
\label{fig:accretion_winds}
\end{figure}

In Figure~\ref{fig:accretion_winds} we show an updated comparison between the
models of \citet{badenes07} and the Fe K$\alpha$ centroids measured for Type Ia
SNRs by \citet{yamaguchi14} (see Section 3 below for a more detailed
explanation of these measurements). The physical radii and the ionization state
of Fe in Type Ia SNRs are in good agreement with SNR models where Type Ia SN
ejecta interact with an undisturbed ambient medium (shaded regions). Models
with a $\rho_{\mathrm{CSM}}$ $\sim$ $Ar^{-2}$ circumstellar profile (blue plot)
lead to small, highly ionized SNRs that do not appear in the observed sample
(though many core-collapse SNRs fit this description, see Section 3). Models
expanding into low density cavities excavated by accretion winds lead to large,
low-ionization SNRs. Although these objects are not common, there is at least
one example in the sample from \citet{yamaguchi14}: SNR RCW 86. This SNR has
long been suspected to be a cavity explosion, and its bulk dynamics are hard to
interpret without a fast, sustained outflow from the progenitor, with
properties similar to those of accretion winds \citep{badenes07,williams11}.
  
\begin{figure}
\includegraphics[width=0.5\textwidth]{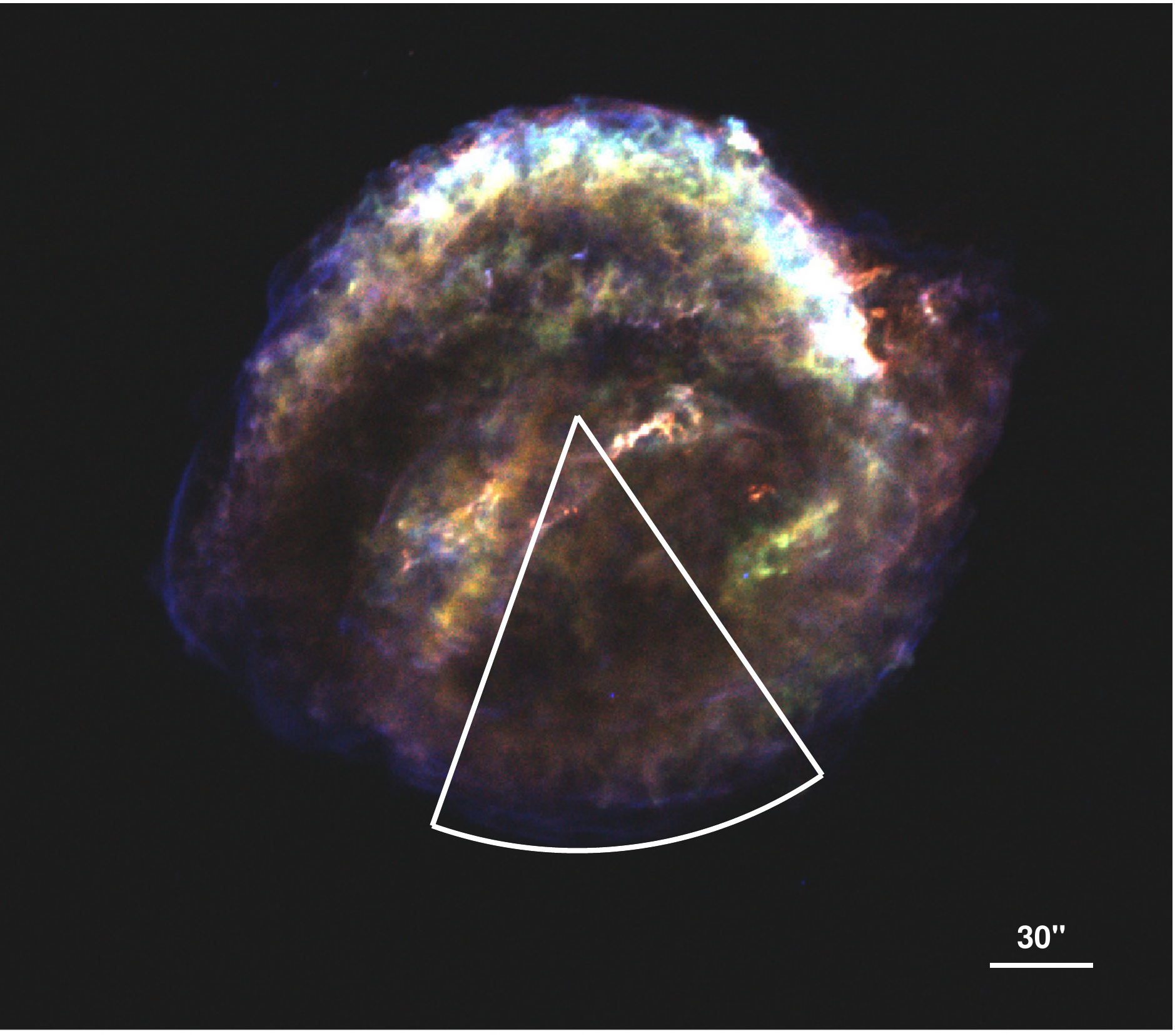}
\includegraphics[width=0.5\textwidth]{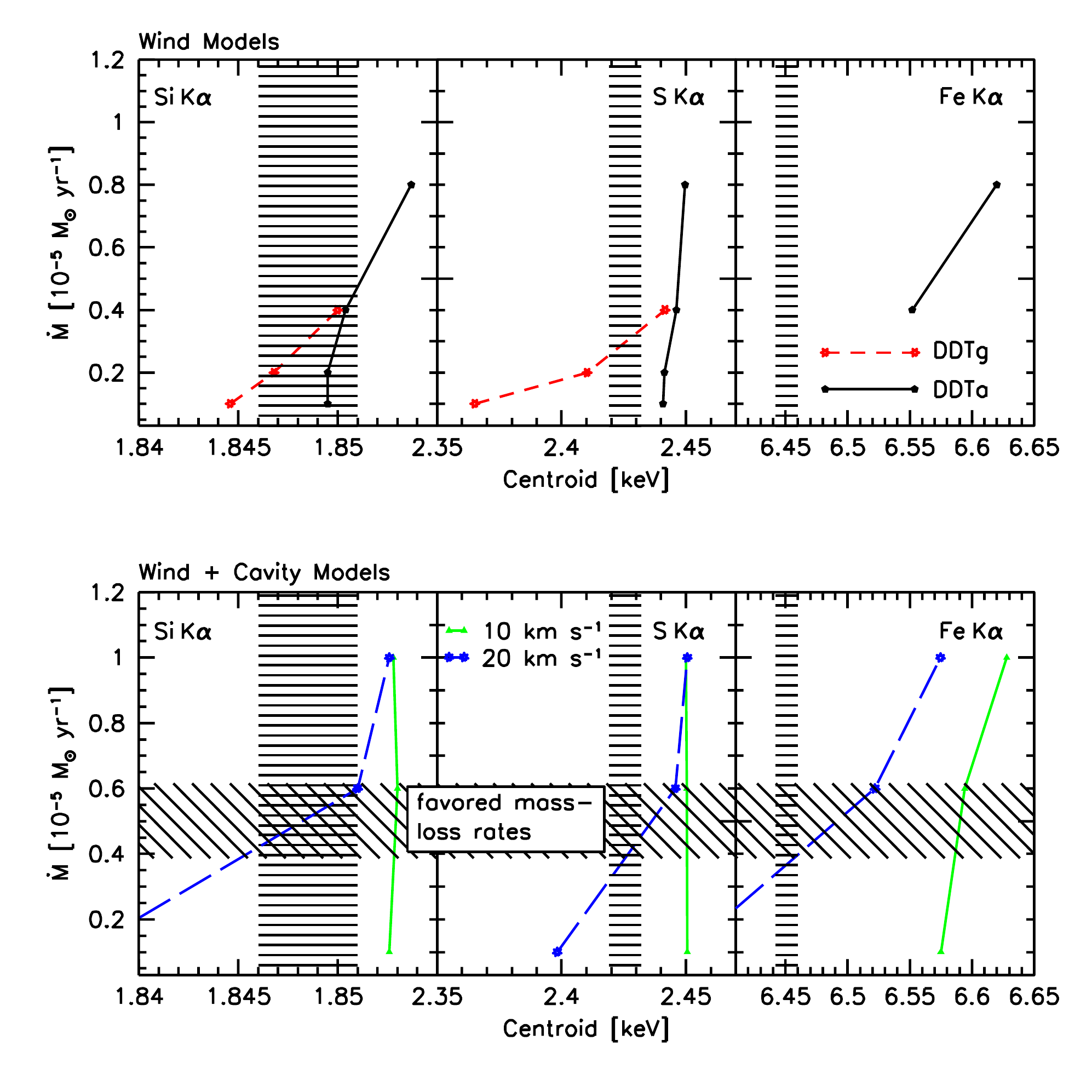}
\caption{{\it Left}: Kepler's SNR viewed in X-rays with {\it Chandra}
ACIS-S3. The RGB image shows $0.4$--$0.75$ keV emission in red, $0.75$--$1.2$ keV
emission in green, and $1.2$--$7.0$ keV emission in blue. 
In the image, north is up and east is to the left. The outlined region
corresponds to the blastwave--ejecta interaction discussed in 
\citet{patnaude12}. 
{\it Right:} In the upper panel, we plot the measured
versus computed line centroids for Si K$\alpha$, S K$\alpha$, and Fe K$\alpha$, 
for the DDTa (black solid) and DDTg (red dashed) models. The hatched region in each panel
corresponds to the measured centroid including the 90\% confidence interval
The DDTg models and a subset of the DDTa models ($\dot{M}$ $<$ 4$\times$10$^{-6}$ 
M$_{\sun}$ yr$^{-1}$) do not produce any 
Fe K emission. In the lower panel, we plot the line centroids for the DDTa
cavity models for v$_{\mathrm{wind}}$ = 10 (green solid) and 20 (blue dashed) km 
s$^{-1}$, for a range of mass-loss rates. The line centroids and errors are indicated by 
the vertical hatched regions in each panel. The allowed mass-loss rates as
dictated by the comparison between the measured and modeled line centroids are 
marked by the horizontal cross hatched region \citep[Figures
reproduced with the permission of the authors;][]{patnaude12}.}
\label{fig:kepler}
\end{figure}

A curious feature of Figure~\ref{fig:accretion_winds} is the agreement shown
between Kepler's SNR and models for an unmodified ambient medium. Shown in
Figure~\ref{fig:kepler} (left) is an X-ray image of Kepler, showing clear signs
of circumstellar interaction. \citet{patnaude12} modeled the dynamics and X-ray
emission for a Type Ia model expanding in the dense wind of an AGB companion
\citep{chiotellis12}. \citet{patnaude12} jointly modeled the blastwave dynamics
and X-ray spectrum of the SNR. The line energy centroids of prominent emission
lines of silicon and iron are sensitive to the circumstellar environment the
remnant is expanding into. \citet{patnaude12} found that in order to match both
the dynamics and bulk X-ray spectral properties, a small cavity, of order 0.1
pc in radius, is required around the progenitor. They suggest that the cavity
could have been formed just prior to thermonuclear runaway, as the accretion
wind pushes out against the dense wind from the AGB companion.

Given the high Fe content observed in the spectrum of Kepler's SNR
\citep{patnaude12,katsuda15a}, there is little doubt that it is the result of a
Ia and the CSM wind scenario argued by
\citet{chiotellis12,patnaude12,katsuda15a} seems to favor the SD channel,
though fast accretion winds appear ruled out. However, \citet{kerzendorf14}
obtained spectra for 24 stars with $L>10L_{\sun}$, and none showed evidence for
being a former donor star, including no red giant, AGB, or post-AGB surviving
companions in the inner 40$\arcsec$ of the remnant. Similar searches for
companions in the LMC remnants SNR 0519-67.5 and 0509-67.5 also turn up no
evidence for a surviving companion \citep{schaefer12,edwards12}. While this
isn't firm proof for the viability of the DD scenario, it does suggest that, at
least in these cases, the double degenerate progenitor model could be required
in order to produce these SNRs. 

To conclude, observations of Type Ia SNRs provide clear evidence for a
M$_{\mathrm{Ch}}$ SN Ia progenitor in SNR 3C 397 \citep{yamaguchi15}, and the
presence of a fast, sustained outflow in SNR RCW 86
\citep{badenes07,williams11}. Taken at face value, this seems to support the SD
scenario for at least some local Type Ia SNe. However, the bulk properties of
most Type Ia SNRs are at odds with the CSM configurations expected in SD
progenitors \citep[see Figure~\ref{fig:accretion_winds} and][for a detailed
discussion]{badenes07}. More contrived CSM profiles are certainly possible in
some cases, as shown by \citet{patnaude12} for Kepler, but a more likely
explanation is that DD progenitors also make a significant contribution to the
local SN Ia rate \citep[see][]{badenes12}. A mixture of progenitor channels has
interesting implications for cosmology \citep{ponder16}.

\subsection{Core Collapse SNRs -- Progenitor Connections}
\label{sec:ccsne}
\index{Core Collapse SNRs}

A good review concerning the connections between young supernova remnants
and their progenitors can be found in \citet{chevalier05}. They note that
different supernova types arise form different zero age main sequence
masses, with SNe IIP occuring in the 8-15 M$_{\sun}$ range, SNe Ib/c
occuring in stars with ZAMS mass $\gtrsim$ 35 M$_{\sun}$, and IIb/IIL
SNe resulting from stars in the middle, with observational results 
providing information on the progenitor to supernova type relationship
\citet[e.g.,][]{smartt15}. 

If firm connections between progenitor and supernova type can be
established, then connecting a remnant to it's parent supernova
provides an indirect path back to the progenitor type. \citet{chevalier05}
attempted this for several objects, including the Crab, 3C~58, and
G292.0+1.8. In the Crab, measured abundances suggest
that the progenitor lacked an O-rich mantle, placing the progenitor
star at the low end of the supernova progenitor mass range, $\sim$
8--10 M$_{\sun}$, suggesting that the Crab supernova was of the 
Type IIP variety. On the other hand, the SNR G292.0+1.8 probably
arises from a more massive progenitor, as the blastwave has swept
up $\sim$ 15--40 M$_{\sun}$ of RSG wind \citep{lee10}.  
Estimates of a high swept up mass points to a massive progenitor, with
an initial mass of 20--35 M$_{\sun}$. 

In general, the range of progenitor models which lead to core collapse
supernova remnants is much larger than one sees in the case of 
Type Ia SNRs. This is reflected in the morphological and compositional 
diversity of CC SNR we observe
in the Galaxy. Ranges in mass loss rates and wind speeds, combined with
the initial mass of the star (which will influence the final abundances),
as well as evolutionary changes that the progenitor may go through 
(such as a Wolf-Rayet phase), not to mention the effects that a binary
companion can have on the progenitor's evolution, leads to larger 
parameter space to explore when comparing CC SNR to their progenitors.
As seen in \citet{lee10} and others, examples exist where broad
statements about the progenitor can be made, but few objects allow for
the same types of detailed analyses that we find in Ia studies. The
exceptions are the Galactic SNR Cassiopeia A (Cas A), and the
youngest know remnant, SN~1987A, in the Large Magallenic Cloud.

\begin{figure}
\includegraphics[width=0.5\textwidth]{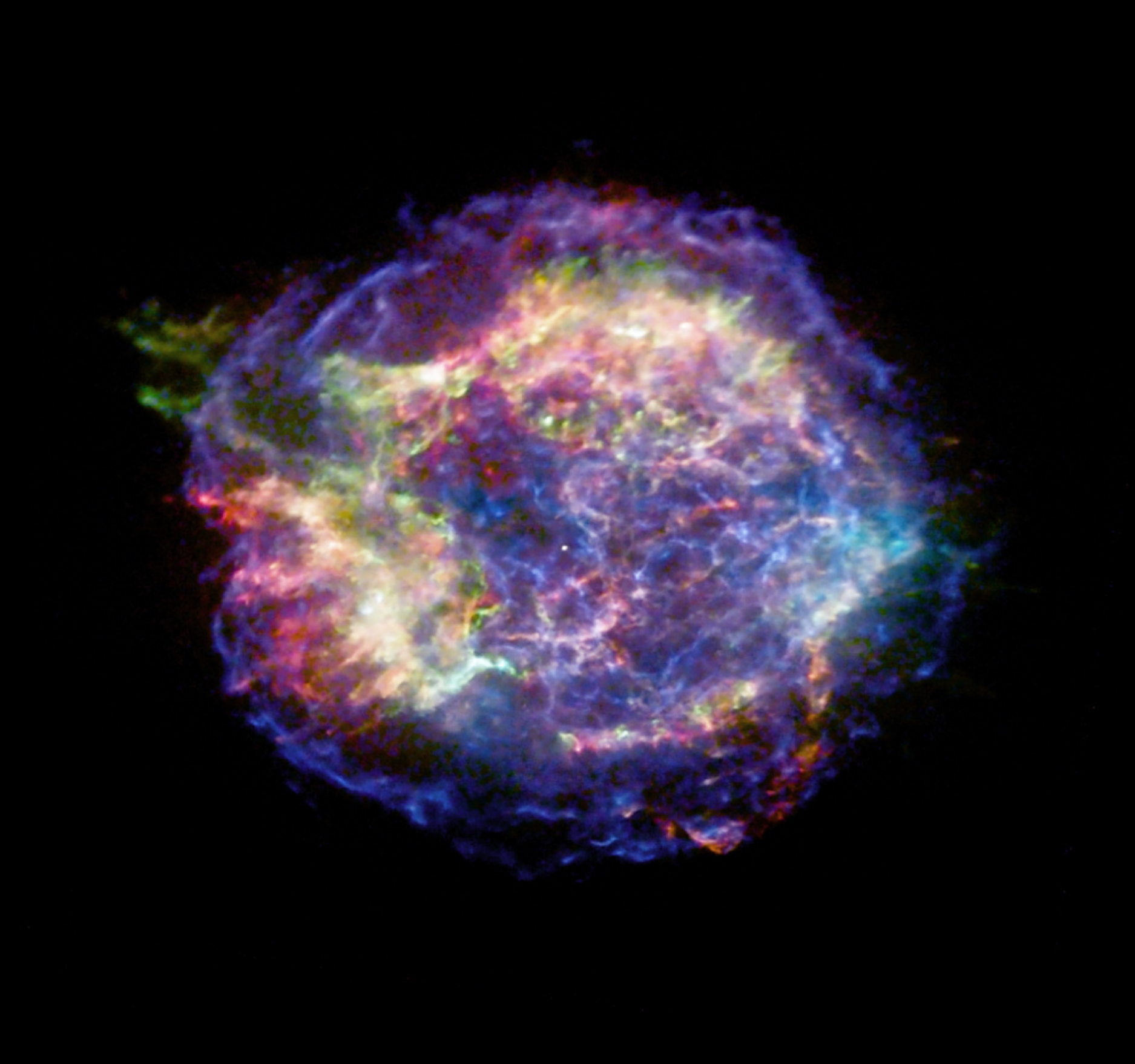}
\includegraphics[width=0.5\textwidth,clip=true]{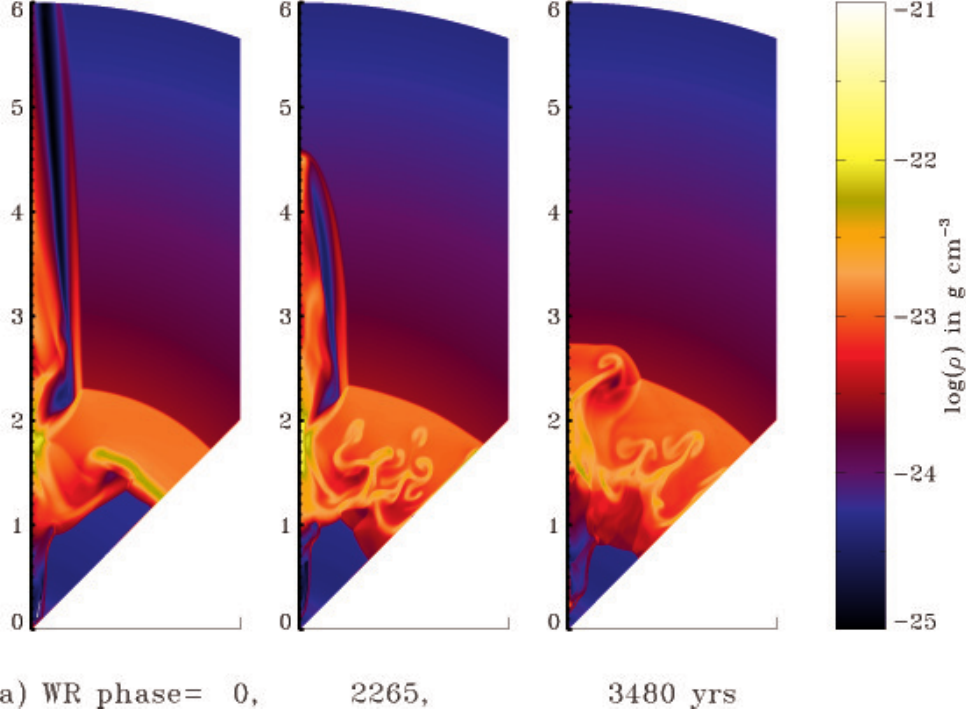}
\caption{{\it Left}: Cas A in X-rays. The three color image shows 0.5--1.5
keV in red, 1.5--3.0 keV in green, and 4.2--6.0 keV in green. The neutron
star is clearly visible near the center. {\it Right}: Density of the 
supernova remnant at a time of 330~yr after explosion. The left panel shows 
the remnant that results from the evolution of ejecta in a CSM 
modified by a pure RSG wind, with the middle and righthand panels
showing the case where the Wolf-Rayet phase lasted for 2265 and 
3480 years, respectively. The forward shock of the remnant in all three cases is 
located at a distance of $\sim 2.4$~pc, while the reverse shock is 
located increasingly farther inward for longer WR phases Details of the models
may be found in \citet{schure08}. 
\citep[Righthand figure reproduced with the permission of the authors;][Figure 4, pp 403]{schure08}.}
\label{fig:casa}
\end{figure}

Cas A, shown in Figure~\ref{fig:casa} (left), may represent the most 
well studied of the Galactic SNR. At an age
of 330 yr \citep{thorstensen01}, it has a morphology 
which is consistent with that of a SNR interacting with a RSG wind
\citep{chevalier03}. \citet{lee14,hwang12} undertook detailed studies
of the mass-loss history of the Cas A progenitor, and estimate that the
progenitor star shed $\gtrsim$ 6M$_{\sun}$ of material prior to 
core--collapse. Combined with the estimated mass of the neutron star
and ejecta, the progenitor zero-age main sequence mass is $\approx$
12 M$_{\sun}$. 

The light echo spectra of the Cas A SN obtained by 
\citet{krause08,rest11} suggest a Type IIb origin, similar to SN~1993J,
implying a red supergiant origin. The analysis by \citet{lee14} suggest
that the Cas A progenitor lost much of its mass through the RSG wind, 
compatible with the IIb SN type, 
though this is at odds with the mass-loss rates adopted for
single RSG stars \citep{woosley02}. Solutions to the high amount of mass
lost from Cas A include episodic mass loss through pulsational 
instabilities in the RSG phase \cite{yoon10}, or enhanced mass loss
through the interaction of a binary companion \citep{young06}.

Another possibility is that the Cas A progenitor went through a short
Wolf-Rayet phase prior to core collapse. The fast wind from a Wolf-Rayet
progenitor would clear out a small cavity, much like is required in 
Kepler's SNR,
and would imprint itself on the dynamics and X-ray emission of the
remnant. \citet{hwang12} required such a cavity in order to match the
observed X-ray emission. \citet{schure08} modeled the evolution of the Cas A
jet in a cavity, and found that if a Wolf-Rayet phase existed, it would be
rather short ($\sim$ a few thousand years; Fig.~\ref{fig:casa}, right). 

Besides Cas A, perhaps the most well studied supernova remnant
is that of SN~1987A, a recent review of which can be found in 
\citet{mccray16}. SN~1987A is the closest example of a remnant where
the progenitor is known through pre-explosion imaging. However, supernovae
are now regularly identified with their progenitors through the use
of pre-explosion images. \citet{smartt15} notes 18 detections of 
supernova progenitors, with 27 additional upper limits -- a large number
of the sample of detected supernova progenitors result in SN IIP, IIL, 
or IIb. This is consistent with \citet{chevalier05}, who associated
many Galactic SNR with Type II SNe, resulting from RSG progenitors of
various masses. Only one or possibly two Galactic SNR have been firmly
associated with SNe Type Ib/c (and hence with a Wolf-Rayet progenitor) --
W49B as a possible gamma-ray burst remnant \citep{lopez13}, and
RX J1713--3946 \citep{katsuda15b}.

\section{SNR Bulk Properties}
\label{sec:bulk}

As discussed in the previous sections, typing individual supernova remnants
remains challenging. Recently. \citet{yamaguchi14} presented a method of typing
SNRs based on the Fe-K$\alpha$ line centroid and luminosity. Since Fe is
produced in the center of the progenitor during the explosion, heating of Fe
can be delayed, resulting in ionization states lower than He-like (Fe$^{24+}$)
in young and middle-aged SNRs. The ionization state affects the Fe-K line
centroid, which can be measured with high precision with satellites such as
{\it Chandra}, {\it XMM--Newton}, or {\it Suzaku}. In brief, they found a
correlation between supernova type, and the Fe-K line centroid: the line
centroid for Type Ia SNRs are generally lower ($<$ 6550 eV) than those found in
CCSNRs. Perhaps more importantly, when they computed synthetic line centroids
from models for Type Ia ejecta interacting with a uniform medium, the models
predict bulk properties in line with the observations.

\citet{patnaude15} extended this work to study core collapse SNR
progenitors. While \citet{yamaguchi14} focussed on Type Ia SNR
in an unmodified circumstellar environment, a key question, as demonstrated
in previous sections, is the nature of the circumstellar
environment around massive progenitors. Red and Yellow supergiants
(R/YSGs) expel several solar masses of material over their lifetimes,
with mass loss rates $\approx$ 10$^{-5}$ M$_{\sun}$ yr$^{-1}$. Wolf-Rayet
stars, thought to be the progenitors of the Ib/c subset of core collapse
supernovae, expel mass at similar rates, but with wind velocities
two orders of magnitude higher. 

\begin{figure}
\includegraphics[width=0.5\textwidth]{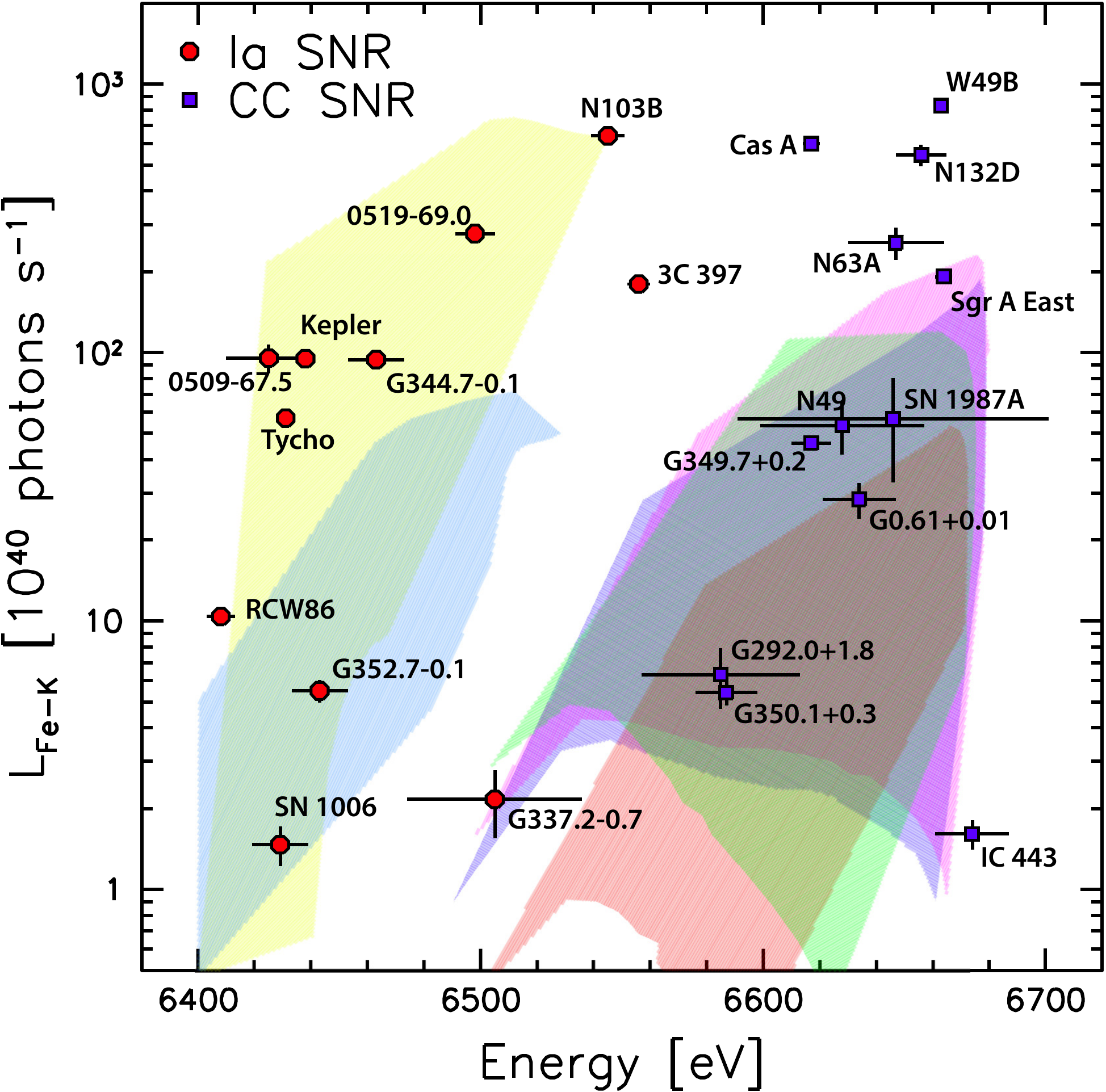}
\includegraphics[width=0.5\textwidth]{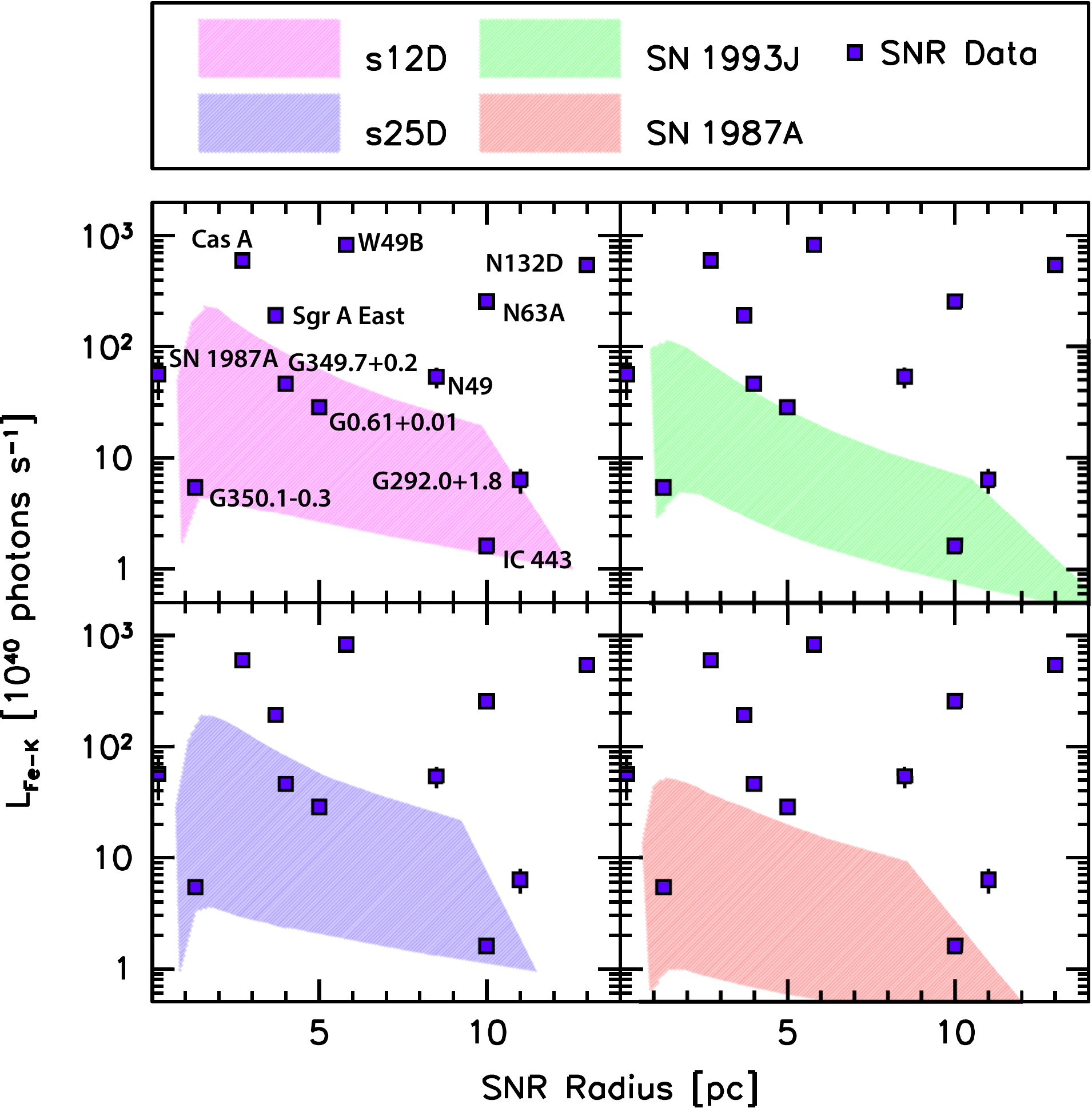}
\caption{{\it Left}: Fe-K line luminosity vs. centroid energy
for Galactic and Magallenic Cloud Ia and core collapse 
supernova remnants measured by \citet{yamaguchi14}. The transparent shaded regions correspond to models
for either Ia or CCSNe ejecta, with yellow and light blue corresponding
to the Ia models DDTa and DDTg, and red, green, magenta, and dark
blue corresponding to CCSNe models SN~1987A, SN~1993J, and a 12 and 
25M$_{\sun}$ solar metallicity progenitor, respectively. The data are
from \citet{yamaguchi14}. {\it Right}: Fe-K line luminosity as a function
of radius for core collapse supernova remnant measurements and models 
\citep[Figures reproduced with the permission of the authors;][]{patnaude15}.}
\label{fig:bulk}
\end{figure}

In addition to steady line-driven
mass loss, episodic mass loss may also be relevant. 
\citet{patnaude15} modeled the circumstellar
environments of core collapse progenitors for a range of mass loss
rates and wind velocities, but did not consider the effects that
episodic mass loss would have on their results.
In Figure~\ref{fig:bulk} (left), we reproduce the results from
\citet{patnaude15}. As seen in the left hand panel, there is a clear
bifurcation between Type Ia and CCSNe ejecta models, consistent with
the observed properties of the remnants. In particular, the CCSNe
models consistently produce higher Fe-K line centroids. \citet{patnaude15}
attribute this to the blastwave interacting with a great deal of
material deposited by the stellar wind around the progenitor.

While these models capture the general spectral
characteristics of the data, when correlating modeled radii to line
luminosity, an apparent disconnect is seen between the modeled line
luminosities as a function of radius, as compared to the data. This
is shown in Figure~\ref{fig:bulk} (right), where the
models underestimate the Fe-K line luminosity compared
to the data. 

In the case of CCSNe, the Fe-K line centroid is already
close to the collisional ionization equilibrium value. Modest increases
in the progenitor mass loss rate will have a small effect on the
ionization, while providing a boost to the line luminosity, since
the X-ray emission scales like the square of the density. 
Increases in the mass loss rate, if non-steady, would
not affect the radius, as the interaction timescale between the blast wave and 
the additional circumstellar material would be short, compared to the
overall age of the remnant.

There is now growing evidence that core collapse progenitors undergo
some sort of episodic mass loss prior to core collapse. The optical light
curves of several Type IIn supernovae, including SN~2005ip, SN~2005kj, and
SN~2006aa, all suggest episodic mass loss prior to core collapse
\citep{moriya14}, and a mass loss
event from the progenitor of SN~2009ip was sufficiently luminous that it 
was misclassified as a SN, though the progenitor did finally explode in 2012. 
The mechanism for the episodic mass loss is not currently understood, but
the mass ejection may be related to wave-driven mass loss during core neon
and oxygen burning \citep{shiode14}, or possibly unsteady nuclear 
burning in the 
envelope \citep{smith14}. 

The mass loss history, steady or otherwise, will imprint itself on the X-ray
emission from the SNR \citep{patnaude15}. As an  example of this,
\citet{moriya12} investigated the progenitors for so-called recombining
supernova remnants, remnants where the ion temperature is sufficiently higher
than the electron temperature, and found that a plausible explanation exists
for them if their progenitors lost a sufficient amount of mass that was
subsequently deposited near the progenitor, just prior to core-collapse. They
argued that the progenitors of these remnants are likely red supergiants, and
are likely not associated with SNe IIn.

\section{Conclusions}
\label{sec:conc}

In this chapter we have reviewed several properties of supernova remnants
which provide links back to their progenitor systems. X-ray observations
of supernova remnants place constraints on the metallicity of the progenitor and 
measuring the explosion energetics, either through proper motion analyses
or inferred from spectral measurements can provide constraints on the 
explosion energetics, as well as the density of the circumstellar 
medium. These parameters relate directly back to the evolution of the
progenitor.

In the case of Type Ia SNRs, measuring the abundances of neutron-rich material
places constraints on the central density of the white
dwarf progenitor. Inferring a high central density points to a M$_{\rm Ch}$
explosion and likely a single-degenerate progenitor system. Likewise, joint
spectral/dynamical fits to objects such as Kepler's SNR point to complicated
evolutionary histories for the progenitor, where an AGB wind is required from
the progenitor system, but only if it has a cavity excised out of it prior to
the supernova event. A similar result has been suggested in Tycho
\citep{chiotellis13,slane14}. 

For progenitors od core-collapse SNR, the situation is more
complicated. CCSNe progenitors exhibit a larger mass range, with order of
magnitude variations in mass loss rates and progenitor wind velocities.  Except for
the notable cases of SN~1987A and Cas A, only broad statements about the SNR
progenitor may be made at this time. However, when one begins to step back from individual
remnants, trends in SNR bulk properties begin to emerge.  Recent
results by \citet{yamaguchi14} show a clear bifurcation in line centroids
between SNR Ia and core-collapse SNRs. These results are consistent with model predictions.

While models for SNR progenitors do generally agree with observed bulk
properties of SNR, there are still inconsistencies. Nonetheless, these issues may be
addressed by stellar evolution models which better define the end stages
of the progenitor, including such affects as episodic mass loss and non-steady
nuclear burning. Both mechanisms alter the circumstellar environment
in a way that is readily observable in remnants, hundreds of years after
the supernova event.

\section*{Acknowledgments}

The authors wish to thank the editorial staff for their steadfast 
patience during
the preparation of this manuscript. D.~J.~P. acknowledges support from 
NASA contract NAS8-03060 and NASA/TM6-17003X. 
C.~B. acknowledges partial support from grants NSF/AST-1412980 and 
NASA/NNX15AM03G-S01.

\section*{Cross References}

\begin{itemize}

\item Supernova Remnant Cassiopeia A

\item Observational Classification of Supernovae

\item Dynamical Evolution and Radiative Processes in Supernova Remnants

\item Supernova remnant from SN~1987A

\end{itemize}

\printindex


\begin{thebibliography}{}
\expandafter\ifx\csname natexlab\endcsname\relax\def\natexlab#1{#1}\fi

\bibitem[Badenes et al.(2003)]{badenes03} Badenes, C., Bravo, E., Borkowski, K.~J., \& Dom{\'{\i}}nguez, I.\ 2003, ``Thermal X-Ray Emission from Shocked Ejecta in Type Ia Supernova Remnants: Prospects for Explosion Mechanism Identification,'' \apj, 593, 358 

\bibitem[Badenes et al.(2005)]{badenes05} Badenes, C., Borkowski, K.~J., \& Bravo, E.\ 2005, ``Thermal X-Ray Emission from Shocked Ejecta in Type Ia Supernova Remnants. II. Parameters Affecting the Spectrum,'' \apj, 624, 198 

\bibitem[Badenes et al.(2006)]{badenes06} Badenes, C., Borkowski, K.~J., Hughes, J.~P., Hwang, U., \& Bravo, E.\ 2006, ``Constraints on the Physics of Type Ia Supernovae from the X-Ray Spectrum of the Tycho Supernova Remnant,'' \apj, 645, 1373 

\bibitem[Badenes et al.(2007)]{badenes07} Badenes, C., Hughes, J.~P., Bravo, E., \& Langer, N.\ 2007, ``Are the Models for Type Ia Supernova Progenitors Consistent with the Properties of Supernova Remnants?,'' \apj, 662, 472 

\bibitem[Badenes et al.(2009)]{badenes09} Badenes, C., Harris, J., Zaritsky, D., \& Prieto, J.~L.\ 2009, ``The Stellar Ancestry of
  Supernovae in the Magellanic Clouds. I. The Most Recent Supernovae in the Large Magellanic Cloud,'' \apj, 700, 727

\bibitem[Badenes(2010)]{badenes10} Badenes, C.\ 2010, ``X-ray studies of supernova remnants: A different view of supernova
  explosions,'' Proceedings of the National Academy of Science, 107, 7141 

\bibitem[Badenes \& Maoz(2012)]{badenes12} Badenes, C., \& Maoz, D.\ 2012, ``The Merger Rate of Binary White Dwarfs in the Galactic Disk,'' \apjl, 749, L11 

\bibitem[Borkowski et al.(2001)]{borkowski01} Borkowski, K.~J., Lyerly, W.~J., \& Reynolds, S.~P.\ 2001, ``Supernova Remnants in
  the Sedov Expansion Phase: Thermal X-Ray Emission,'' \apj, 548, 820 

\bibitem[Borkowski et al.(2014)]{borkowski14} Borkowski, K.~J., Reynolds, S.~P., Green, D.~A., et al.\ 2014, ``Nonuniform Expansion of the Youngest Galactic Supernova Remnant G1.9+0.3,'' \apjl, 790, L18 

\bibitem[Brachwitz et al.(2000)]{brachwitz00} Brachwitz, F., Dean, D.~J., Hix, W.~R., et al.\ 2000, ``The Role of Electron Captures in Chandrasekhar-Mass Models for Type IA Supernovae,'' \apj, 536, 934 

\bibitem[Chevalier \& Oishi(2003)]{chevalier03} Chevalier, R.~A., \& Oishi, J.\ 2003, ``Cassiopeia A and Its Clumpy Presupernova Wind,'' \apjl, 593, L23 

\bibitem[Chevalier(2005)]{chevalier05} Chevalier, R.~A.\ 2005, ``Young Core-Collapse Supernova Remnants and Their Supernovae,'' \apj, 619, 839 

\bibitem[Chiotellis et al.(2012)]{chiotellis12} Chiotellis, A., Schure, K.~M., \& Vink, J.\ 2012, ``The imprint of a symbiotic binary progenitor on the properties of Kepler's supernova remnant,'' \aap, 537, A139 

\bibitem[Chiotellis et al.(2013)]{chiotellis13} Chiotellis, A., Kosenko, D., Schure, K.~M., Vink, J., \& Kaastra, J.~S.\ 2013, ``Modelling the interaction of thermonuclear supernova remnants with circumstellar structures: the case of Tycho's supernova remnant,'' \mnras, 435, 1659 

\bibitem[Chomiuk et al.(2016)]{chomiuk16} Chomiuk, L., Soderberg, A.~M., Chevalier, R.~A., et al.\ 2016, ``A Deep Search for Prompt Radio Emission from Thermonuclear Supernovae with the Very Large Array,'' \apj, 821, 119 

\bibitem[DeLaney \& Rudnick(2003)]{delaney03} DeLaney, T., \& Rudnick, L.\ 2003, ``The First Measurement of Cassiopeia A's Forward Shock Expansion Rate,'' \apj, 589, 818 

\bibitem[Dessart et al.(2014)]{dessart14} Dessart, L., Blondin, S., Hillier, D.~J., \& Khokhlov, A.\ 2014, ``Constraints on the explosion mechanism and progenitors of Type Ia supernovae,'' \mnras, 441, 532 

\bibitem[Edwards et al.(2012)]{edwards12} Edwards, Z.~I., Pagnotta, A., \& Schaefer, B.~E.\ 2012, ``The Progenitor of the Type Ia Supernova that Created SNR 0519-69.0 in the Large Magellanic Cloud,'' \apjl, 747, L19 

\bibitem[Ferri{\`e}re(2001)]{ferriere01} Ferri{\`e}re, K.~M.\ 2001, ``The interstellar environment of our galaxy,'' Reviews of Modern Physics, 73, 1031 


\bibitem[Filippenko(1997)]{filippenko97} Filippenko, A.~V.\ 1997, ``Optical
Spectra of Supernovae,'' \araa, 35, 309 

\bibitem[Hachisu et al.(1996)]{hachisu96} Hachisu, I., Kato, M., \& Nomoto, K.\ 1996, ``A New Model for Progenitor Systems of Type IA Supernovae,'' \apjl, 470, L97 


\bibitem[Hillebrandt et al.(2013)]{hillebrandt13} Hillebrandt, W., Kromer, M., R{\"o}pke, F.~K., \& Ruiter, A.~J.\ 2013, ``Towards an understanding of Type Ia supernovae from a synthesis of theory and observations,'' Frontiers of Physics, 8, 116 


\bibitem[Hwang \& Laming(2009)]{hwang09} Hwang, U., \& Laming, J.~M.\ 2009, ``The Circumstellar Medium of Cassiopeia a Inferred from the Outer Ejecta Knot Properties,'' \apj, 703, 883 


\bibitem[Hwang \& Laming(2012)]{hwang12} Hwang, U., \& Laming, J.~M.\ 2012, ``A Chandra X-Ray Survey of Ejecta in the Cassiopeia A Supernova Remnant,'' \apj, 746, 130

\bibitem[Jennings et al.(2014)]{jennings14} Jennings, Z.~G., Williams, B.~F., Murphy, J.~W., et al.\ 2014, ``The Supernova Progenitor Mass Distributions of M31 and M33: Further Evidence for an Upper Mass Limit,'' \apj, 795, 170

\bibitem[Katsuda et al.(2015a)]{katsuda15a} Katsuda, S., Mori, K., Maeda, K., et al.\ 2015, ``Kepler's Supernova: An Overluminous Type Ia Event Interacting with a Massive Circumstellar Medium at a Very Late Phase,'' \apj, 808, 49 

\bibitem[Katsuda et al.(2015b)]{katsuda15b} Katsuda, S., Acero, F., Tominaga, N., et al.\ 2015, \apj, 814, 29 

\bibitem[Kerzendorf et al.(2014)]{kerzendorf14} Kerzendorf, W.~E., Childress, M., Scharw{\"a}chter, J., Do, T., \& Schmidt, B.~P.\ 2014, ``A Reconnaissance of the Possible Donor Stars to the Kepler Supernova,'' \apj, 782, 27 

\bibitem[Koo \& McKee(1992)]{koo92} Koo, B.-C., \& McKee, C.~F.\ 1992, ``Dynamics of wind bubbles and superbubbles. I - Slow winds and fast winds. II - Analytic theory,'' \apj, 388, 93 

\bibitem[Krause et al.(2008)]{krause08} Krause, O., Birkmann, S.~M., Usuda, T., et al.\ 2008, ``The Cassiopeia A Supernova Was of Type IIb,'' Science, 320, 1195

bibitem[Kerzendorf et al.(2014)]{kerzendorf14} Kerzendorf, W.~E., Childress, M., Scharw{\"a}chter, J., Do, T., \& Schmidt, B.~P.\ 2014, ``A Reconnaissance of the Possible Donor Stars to the Kepler Supernova,'' \apj, 782, 27 

\bibitem[Lee et al.(2010)]{lee10} Lee, J.-J., Park, S., Hughes, J.~P., et al.\ 2010, ``The Outer Shock of the Oxygen-Rich Supernova Remnant G292.0+1.8: Evidence for the Interaction with the Stellar Winds from Its Massive Progenitor,'' \apj, 711, 861 

\bibitem[Lee et al.(2014)]{lee14} Lee, J.-J., Park, S., Hughes, J.~P., \& Slane, P.~O.\ 2014, ``X-Ray Observation of the Shocked Red Supergiant Wind of Cassiopeia A,'' \apj, 789, 7 

\bibitem[Lopez et al.(2013)]{lopez13} Lopez, L.~A., Ramirez-Ruiz, E., Castro, D., \& Pearson, S.\ 2013, ``The Galactic Supernova Remnant W49B Likely Originates from a Jet-driven, Core-collapse Explosion,'' \apj, 764, 50 

\bibitem[Maggi et al.(2016)]{maggi16} Maggi, P., Haberl, F., Kavanagh, P.~J., et al.\ 2016, ``The population of X-ray supernova remnants in the Large Magellanic Cloud,'' \aap, 585, A162 

\bibitem[Maoz et al.(2014)]{maoz14} Maoz, D., Mannucci, F., \& Nelemans, G.\ 2014, ``Observational Clues to the Progenitors of Type Ia Supernovae,'' \araa, 52, 107 

\bibitem[Margutti et al.(2014)]{margutti14} Margutti, R., Parrent, J., Kamble, A., et al.\ 2014, ``No X-Rays from the Very Nearby Type Ia SN 2014J: Constraints on Its Environment,'' \apj, 790, 52 

\bibitem[Maund et al.(2004)]{maund04} Maund, J.~R., Smartt, S.~J., Kudritzki, R.~P., Podsiadlowski, P., \& Gilmore, G.~F.\ 2004, ``The massive binary companion star to the progenitor of supernova 1993J,'' \nat, 427, 129 

\bibitem[McCray \& Fransson(2016)]{mccray16} McCray, R., \& Fransson, C.\ 2016, ``The Remnant of Supernova 1987A,'' \araa, 54, 19 

\bibitem[Moriya(2012)]{moriya12} Moriya, T.~J.\ 2012, ``Progenitors of Recombining Supernova Remnants,'' \apjl, 750, L13

\bibitem[Moriya et al.(2014)]{moriya14} Moriya, T.~J., Maeda, K., Taddia, F., et al.\ 2014, ``Mass-loss histories of Type IIn supernova progenitors within decades before their explosion,'' \mnras, 439, 2917

\bibitem[Park et al.(2013)]{park13} Park, S., Badenes, C., Mori, K., et al.\ 2013, ``A Super-solar Metallicity for the Progenitor of Kepler's Supernova,'' \apjl, 767, L10

\bibitem[Patnaude \& Fesen(2009)]{patnaude09} Patnaude, D.~J., \& Fesen, R.~A.\ 2009, ``Proper Motions and Brightness Variations of Nonthermal X-ray Filaments in the Cassiopeia A Supernova Remnant'', \apj, 697, 535 

\bibitem[Patnaude et al.(2012)]{patnaude12} Patnaude, D.~J., Badenes, C., Park, S., \& Laming, J.~M.\ 2012, ``The Origin of Kepler's Supernova Remnant,'' \apj, 756, 6 

\bibitem[Patnaude et al.(2015)]{patnaude15} Patnaude, D.~J., Lee, S.-H., Slane, P.~O., et al.\ 2015, ``Are Models for
  Core-collapse Supernova Progenitors Consistent with the Properties of Supernova Remnants?,'' \apj, 803, 101 

\bibitem[P{\'e}rez-Torres et al.(2014)]{perez14} P{\'e}rez-Torres, M.~A., Lundqvist, P., Beswick, R.~J., et al.\ 2014,
  ``Constraints on the Progenitor System and the Environs of SN 2014J from Deep Radio Observations,'' \apj, 792, 38 

\bibitem[Ponder et al.(2016)]{ponder16} Ponder, K.~A., Wood-Vasey, W.~M., \& Zentner, A.~R.\ 2016, ``Incorporating Astrophysical Systematics into a Generalized Likelihood for Cosmology with Type Ia Supernovae,'' \apj, 825, 35 


\bibitem[Raymond et al.(2007)]{raymond07} Raymond, J.~C., Korreck, K.~E., Sedlacek, Q.~C., et al.\ 2007, ``The Preshock Gas of SN 1006 from Hubble Space Telescope Advanced Camera for Surveys Observations,'' \apj, 659, 1257 


\bibitem[Rest et al.(2008)]{rest08} Rest, A., Welch, D.~L., Suntzeff, N.~B., et al.\ 2008, ``Scattered-Light Echoes from the Historical Galactic Supernovae Cassiopeia A and Tycho (SN 1572),'' \apjl, 681, L81 

\bibitem[Rest et al.(2011)]{rest11} Rest, A., Foley, R.~J., Sinnott, B., et al.\ 2011, ``Direct Confirmation of the Asymmetry of
  the Cas A Supernova with Light Echoes,'' \apj, 732, 3

\bibitem[Saio \& Nomoto(1985)]{saio85} Saio, H., \& Nomoto, K.\ 1985, ``Evolution of a merging pair of C+O white dwarfs to form a single neutron star,'' \aap, 150, L21 

\bibitem[Schaefer \& Pagnotta(2012)]{schaefer12} Schaefer, B.~E., \& Pagnotta, A.\ 2012, ``An absence of ex-companion stars in the type Ia supernova remnant SNR 0509-67.5,'' \nat, 481, 164 

\bibitem[Schure et al.(2008)]{schure08} Schure, K.~M., Vink, J., Garc{\'{\i}}a-Segura, G., \& Achterberg, A.\ 2008, ``Jets as Diagnostics of the Circumstellar Medium and the Explosion Energetics of Supernovae: The Case of Cassiopeia A,'' \apj, 686, 399-407

\bibitem[Shen et al.(2013)]{shen13} Shen, K.~J., Guillochon, J., \& Foley, R.~J.\ 2013, ``Circumstellar Absorption in Double
  Detonation Type Ia Supernovae,'' \apjl, 770, L35 

\bibitem[Shen \& Bildsten(2007)]{shen07} Shen, K.~J., \& Bildsten, L.\ 2007, ``Thermally Stable Nuclear Burning on Accreting White
  Dwarfs,'' \apj, 660, 1444 

\bibitem[Shiode \& Quataert(2014)]{shiode14} Shiode, J.~H., \& Quataert, E.\ 2014, ``Setting the Stage for Circumstellar Interaction in Core-Collapse Supernovae. II. Wave-driven Mass Loss in Supernova Progenitors,'' \apj, 780, 96

\bibitem[Slane et al.(2014)]{slane14} Slane, P., Lee, S.-H., Ellison, D.~C., et al.\ 2014, ``A CR-hydro-NEI Model of the Structure and Broadband Emission from Tycho's Supernova Remnant ,'' \apj, 783, 33 


\bibitem[Smartt et al.(2003)]{smartt03} Smartt, S.~J., Maund, J.~R., Gilmore, G.~F., et al.\ 2003, ``Mass limits for the progenitor star of supernova 2001du and other Type II-P supernovae,'' \mnras, 343, 735 

\bibitem[Smartt(2015)]{smartt15} Smartt, S.~J.\ 2015, ``Observational Constraints on the Progenitors of Core-Collapse Supernovae: The Case for Missing High-Mass Stars,'' Publications of the Astronomical Society of Australia, 32, e016 

\bibitem[Smith(2014)]{smith14} Smith, N.\ 2014, ``Mass Loss: Its Effect on the Evolution and Fate of High-Mass Stars,'' \araa, 52, 487 

\bibitem[Thorstensen et al.(2001)]{thorstensen01} Thorstensen, J.~R., Fesen, R.~A., \& van den Bergh, S.\ 2001, ``The Expansion Center and Dynamical Age of the Galactic Supernova Remnant Cassiopeia A,'' \aj, 122, 297 

\bibitem[van Kerkwijk et al.(2010)]{vanKerkwijk10} van Kerkwijk, M.~H., Chang, P., \& Justham, S.\ 2010, ``Sub-Chandrasekhar White Dwarf Mergers as the Progenitors of Type Ia Supernovae,'' \apjl, 722, L157 

\bibitem[Van Dyk et al.(2002)]{vandyk02} Van Dyk, S.~D., Garnavich, P.~M., Filippenko, A.~V., et al.\ 2002, ``The Progenitor of Supernova 1993J Revisited,'' \pasp, 114, 1322 


\bibitem[Van Dyk et al.(2003)]{vandyk03} Van Dyk, S.~D., Li, W., \& Filippenko, A.~V.\ 2003, ``On the Progenitor of the Type II-Plateau Supernova 2003gd in M74,'' \pasp, 115, 1289 

\bibitem[Vink(2012)]{vink12} Vink, J.\ 2012, ``Supernova remnants: the X-ray perspective,'' \aapr, 20, 49 

\bibitem[Wang \& Han(2012)]{wang12} Wang, B., \& Han, Z.\ 2012, ``Progenitors of type Ia supernovae,'' \nar, 56, 122 

\bibitem[Williams et al.(2011)]{williams11} Williams, B.~J., Blair, W.~P., Blondin, J.~M., et al.\ 2011, ``RCW 86: A Type Ia Supernova in a Wind-blown Bubble,'' \apj, 741, 96 

\bibitem[Williams et al.(2014)]{williams14} Williams, B.~F., Peterson, S., Murphy, J., et al.\ 2014, ``Constraints for the Progenitor Masses of 17 Historic Core-collapse Supernovae,'' \apj, 791, 105

\bibitem[Woosley et al.(2002)]{woosley02} Woosley, S.~E., Heger, A., \& Weaver, T.~A.\ 2002, ``The evolution and explosion of massive stars,'' Reviews of Modern Physics, 74, 1015 

\bibitem[Yamaguchi et al.(2014)]{yamaguchi14} Yamaguchi, H., Badenes, C., Petre, R., et al.\ 2014, ``Discriminating the Progenitor Type of Supernova Remnants with Iron K-shell Emission,'' \apjl, 785, L27 

\bibitem[Yamaguchi et al.(2015)]{yamaguchi15} Yamaguchi, H., Badenes, C., Foster, A.~R., et al.\ 2015, ``A Chandrasekhar Mass Progenitor for the Type Ia Supernova Remnant 3C 397 from the Enhanced Abundances of Nickel and Manganese,'' \apjl, 801, L31 

\bibitem[Yang et al.(2013)]{yang13} Yang, X.~J., Tsunemi, H., Lu, F.~J., et al.\ 2013, ``Cr-K Emission Line as a Constraint on the Progenitor Properties of Supernova Remnants,'' \apj, 766, 44 

\bibitem[Yoon \& Cantiello(2010)]{yoon10} Yoon, S.-C., \& Cantiello, M.\ 2010, ``Evolution of Massive Stars with Pulsation-driven Superwinds During the Red Supergiant Phase,'' \apjl, 717, L62 

\bibitem[Young et al.(2006)]{young06} Young, P.~A., Fryer, C.~L., Hungerford, A., et al.\ 2006, ``Constraints on the Progenitor of Cassiopeia A,'' \apj, 640, 891 

\bibitem[Zhou et al.(2016)]{zhou16} Zhou, P., Chen, Y., Zhang, Z.-Y., et al.\ 2016, ``Expanding Molecular Bubble Surrounding Tycho's Supernova Remnant (SN~1572) Observed with the IRAM 30m Telescope: Evidence for a Single-degenerate Progenitor,'' \apj, 826, 34

\end{thebibliography}
\end{document}